\documentclass[11pt]{article}
\usepackage{amsmath}
\usepackage{amsfonts}
\usepackage{amssymb}
\usepackage{theorem}

\def\Th{\Theta}
\def\l{\lambda}
\def\p{\partial}
\newtheorem{theorem}{Proposition}
\newtheorem{lemma}{Lemma}

\newcommand{\be}{\begin{equation}}
\newcommand{\ee}{\end{equation}}
\newcommand{\bea}{\begin{eqnarray}}
\newcommand{\eea}{\end{eqnarray}}
\newcommand{\beaa}{\begin{eqnarray*}}
\newcommand{\eeaa}{\end{eqnarray*}}

\newcommand{\nn}{\nonumber}
\renewcommand{\d}{\mathrm{d}}
\begin{document}
\title{Dunajski generalization of the second heavenly equation:
dressing method and the hierarchy}
\author{
L.V. Bogdanov\thanks
{L.D. Landau ITP, Kosygin str. 2,
Moscow 119334, Russia, e-mail
leonid@landau.ac.ru},
V.S. Dryuma\thanks
{IMI AS RM, Academy str,5,
MD2001, Kishinev, Moldova},
S.V. Manakov\thanks
{L.D. Landau ITP, Kosygin str. 2,
Moscow 119334, Russia}
}
\maketitle
\begin{abstract}
Dunajski generalization of the second heavenly equation
is studied.
A dressing scheme applicable to Dunajski
equation is developed, an example of constructing solutions
in terms of implicit functions is considered. Dunajski equation hierarchy
is described,
its Lax-Sato form is presented.
Dunajsky equation hierarchy is characterized by
conservation of three-dimensional volume form, in which a spectral
variable is taken into account. 
\end{abstract}
\section{Dunajski equation}
A preliminary sketch of the results  presented here 
was given in the preprint \cite{BDM}.

In this work we study an integrable model proposed by Dunajski \cite{Dun}.
This model is a representative of the class of integrable sytems
arising in the context of complex relativity \cite{Plebanski}-\cite{dun3}.
It is closely connected
to the Pleba\'nski second heavenly equation \cite{Plebanski}
and in some sense generalizes it.
The starting point is a (2,2) signature metric in canonical
Pleba\'nski form 
\be
\label{metric}
g=\d w\d x+\d z \d y-\Th_{xx}\d z^2-\Th_{yy}\d w^2+2\Th_{xy}\d w\d z.
\ee
Vacuum Einstein equations and 
conformal anti-self-duality (ASD) condition for this metric
lead to the selebrated Pleba\'nski 
second heavenly equation \cite{Plebanski},
$$
\Th_{wx}+\Th_{zy}+\Th_{xx}\Th_{yy}-\Th_{xy}^2=0.
$$
This equation
is one most known integrable models of 4D selfdual gravity.

Dunajski has
demonstrated that omitting one of the conditions (Einstein equations)
and imposing only conformal anti-self-duality, one still gets an integrable
system, which can be written as
\be
\label{nk1}
\Th_{wx}+\Th_{zy}+\Th_{xx}\Th_{yy}-\Th_{xy}^2=f,
\ee
\be
\label{nk2}
f_{xw}+f_{yz}+
\Th_{yy}f_{xx}+\Th_{xx}f_{yy}-2\Th_{xy}f_{xy}=0.
\ee

Equations (\ref{nk1}), (\ref{nk2}) represent a compatibility condition
for the linear system $L_0\Psi=L_1\Psi=0$, where
$\Psi=\Psi(w, z, x, y, \l)$
and
\begin{eqnarray}
\label{Lax2}
L_0&=&(\p_w-\Th_{xy}\p_y+\Th_{yy} \p_x)-\l\p_y+f_y\p_{\l},\nonumber\\
L_1&=&(\p_z+\Th_{xx}\p_y-\Th_{xy} \p_x)+\l\p_x-f_x\p_{\l}.
\end{eqnarray}

A similar model, corresponding to the first heavenly equation,
was introduced in \cite{Park}. It was demonstrated that a volume-preserving
Riemann-Hilbert problem is connected with the model \cite{Takasaki0}.

In the context of integrable systems, the place of Dunajsky equation
is in the class of integrable equations connected with 
commutation of vector fields. Important and extensively studied
representatives of this class are 
dispersionless
integrable equations, and also 
heavenly equations and hyper-K\"ahler hierarchies
\cite{Takasaki,Takasaki1}. Dunajski equation 
combines the properties of both cases. Similar to dispersionless 
integrable systems (and opposite to heavenly type equations), 
it contains a derivative over spectral parameter
in the Lax pair (\ref{Lax2}).
Connections of Dunajski equation with the theory of 
ordinary differential equations were considered in \cite{Dryum}.

In this work we study Dunajski 
equation (\ref{nk1}), (\ref{nk2}) as an integrable system. The dressing method
based on nonlinear volume-preserving
Riemann-Hilbert problem is developed and used 
to construct special solutions.
Dunajski equation hierachy is described in terms of volume form, its Lax-Sato 
equations
are presented. Compatibility of the flows of the hierarchy is proved.

\section{Dressing scheme}
Let us consider nonlinear vector  Riemann problem of the form
\bea
\mathbf{\Psi}_+= \mathbf{F}(\mathbf{\Psi}_-),
\label{Riemann}
\eea
where $\mathbf{\Psi}_+$, $\mathbf{\Psi}_-$ denote the boundary values
of the N-component vector function on the sides of some
oriented curve $\gamma$ in the complex plane of the variable
$\lambda$. The problem is to find the function analytic
outside the curve with some fixed behavior at infinity (normalization) 
which satisfies
(\ref{Riemann}). The problem of this type was used by Takasaki 
\cite{Takasaki0,Takasaki,Takasaki1}, who stressed its connection to Penrose nonlinear
graviton construction.  
The  problem (\ref{Riemann}) is connected with a class of integrable 
equations,
which can be represented as a commutation relation for 
vector fields containing
a derivative on the spectral variable (see \cite{MS1,MS2,MS3}).
We give a sketch of the dressing scheme
associated with this problem.

The more specific setting relevant for Dunajski equation is the following.
We consider three-component Riemann problem (\ref{Riemann})
\bea
&&\Psi^0_+=F^0(\Psi^0_-,\Psi^1_-,\Psi^2_-),
\nn\\
&&\Psi^1_+=F^1(\Psi^0_-,\Psi^1_-,\Psi^2_-),
\nn\\
&&\Psi^2_+=F^2(\Psi^0_-,\Psi^1_-,\Psi^2_-)
\label{Riemann0}
\eea
for the functions
\bea
&&
\Psi^0\rightarrow\lambda+O(\frac{1}{\lambda}),\nn
\\&&
\Psi^1\rightarrow -\lambda z+x+O(\frac{1}{\lambda}),\nn
\\&&
\Psi^2\rightarrow \lambda w+y+O(\frac{1}{\lambda}),\quad \l\rightarrow \infty,
\label{asympt}
\eea
where $x,y,w,z$ are the variables of Dunajski equation (`times').
We suggest that for given $\mathbf F$ solution of the problem
($\ref{Riemann0}$) for the functions (\ref{asympt})
exists and is unique (at least locally
in $x,y,w,z$).

Let us consider a linearized problem
\beaa
\delta \Psi^i_+=\sum_{j=0,1,2} F^i_{,j}\delta \Psi^j_-.
\eeaa
The linear space of solutions of this problem is spanned by the functions
$\mathbf{\Psi}_x$, $\mathbf{\Psi}_y$, $\mathbf{\Psi}_\lambda$,
which can be multiplied by arbitrary function of spectral variable and times.
The presence of $\mathbf{\Psi}_\lambda$ in the basis is the main difference
between the dressing schemes for the heavenly equation \cite{MS1,heav} and
Dunajski equation.

Expanding the functions $\mathbf{\Psi}_z$, $\mathbf{\Psi}_w$ into the basis,
we obtain linear equations
\begin{eqnarray}
\label{Lax3}
((\p_w+u_{y}\p_y+v_{y}\p_x)-\l\p_y+f_{y}\p_{\l})\mathbf{\Psi}=0,
\nonumber\\
((\p_z-u_{x}\p_y-v_{x}\p_x)+\l\p_x-f_{x}\p_{\l})\mathbf{\Psi}=0,
\end{eqnarray}
where $u$, $v$, $f$ can be expressed through the coefficients of expansion
of $\Psi^0$, $\Psi^1$, $\Psi^2$
at $\lambda=\infty$,
\bea
&&
u=\Psi^2_1-w\Psi^0_1,\quad v=\Psi^1_1+z\Psi^0_1, \quad f=\Psi^0_1,
\label{uvf}
\\&&
\Psi^0=\lambda+
\sum_{n=1}^\infty\frac{\Psi^0_n}{\lambda^n},\;
\Psi^1=-z\l+x+
\sum_{n=1}^\infty\frac{\Psi^1_n}{\lambda^n},\;
\nn\\&&
\Psi^2=w\l+y+
\sum_{n=1}^\infty\frac{\Psi^2_n}{\lambda^n},\quad \l=\infty.
\nn
\eea
The compatibility condition for equations (\ref{Lax3})
represents a closed system of equations 
for three functions $u,v,f$, and solutions to this system
can be found using the problem (\ref{Riemann0}).

To get a Lax pair for Dunajski equation
from linear equations (\ref{Lax3}), we should consider the reduction
$
v_x=-u_y
$,
then we can introduce a potential $\Th$,
\be
v=\Th_y,\; u=-\Th_x.
\label{uvtheta}
\ee
\begin{theorem}
Sufficient condition to provide the reduction
$$
v_x=-u_y
$$
in terms of the Riemann problem
(\ref{Riemann0}) is
\be
\det F^i_{,j}=1.
\label{Rvol}
\ee
\end{theorem}
\textbf{Proof} Condition (\ref{Rvol}) implies that
$$
\d \Psi^0_+\wedge \d \Psi^1_+\wedge \d \Psi^2_+=\d \Psi^0_-\wedge \d \Psi^1_-\wedge \d \Psi^2_-,
$$
and thus the form
$$
\Omega=\d \Psi^0\wedge \d \Psi^1\wedge \d \Psi^2
$$
is analytic in the complex plane. Then the determinant of the matrix
\be
J=
\begin{pmatrix}
\Psi^0_\l&\Psi^1_\l&\Psi^2_\l\\
\Psi^0_x&\Psi^1_x&\Psi^2_x\\
\Psi^0_y&\Psi^1_y&\Psi^2_y
\end{pmatrix},
\label{J0}
\ee
is also analytic, and, considering behavior of this determinant at $\lambda=\infty$,
we come to the conclusion that
$$
\det J=1.
$$
Calculating the coefficient of expansion of $\det J$ at $\lambda=\infty$
corresponding to $\lambda^{-1}$, we get
$$
\Psi^1_{1x}+z\Psi^0_{1x}+\Psi^2_{1y}-w\Psi^0_{1y}=0,
$$
then, according to (\ref{uvf}),
$$
v_x=-u_y.
$$
\hfill$\square$\\
\subsection{An example}
Now let us consider a simple example of constructing solution to Dunajski equation
using problem (\ref{Riemann0}). We introduce the problem of the form
\bea
&&
\Psi^1_+=\Psi^1_-,
\label{ex1}
\\&&
\Psi^2_+=\Psi^2_-\exp(-\mathrm{i}F(\Psi^2_-\Psi^0_-,\Psi^1_-)),
\label{ex2}
\\&&
\Psi^0_+=\Psi^0_-\exp(\mathrm{i}F(\Psi^2_-\Psi^0_-,\Psi^1_-)),
\label{ex3}
\eea
defined on the real axis (or a segment of it),
where $F$ is an arbitrary function of two variables.

It is easy to check that the reduction condition (\ref{Rvol}) is indeed
satisfied in this case.
Equation (\ref{ex1}) implies that $\Psi^1=-\lambda z+x$. Substituting
this solution to linear
equations (\ref{Lax3}) (or using expression (\ref{uvf})), 
we obtain $v=zf$.

The second important property of the problem (\ref{ex1}), (\ref{ex2}), (\ref{ex3})
we use is that $\Psi^2_-\Psi^0_-=\Psi^2_+\Psi^0_+$, thus
the function
$\phi=\Psi^2\Psi^0$ is analytic. Then, taking into account behaviour of functions
at infinity (\ref{asympt}), we come to the conclusion that $\phi$ is
a polynomial of the form
\be
\phi=\Psi^2\Psi^0=\l^2 w+\l y+2fw +u.
\label{phi}
\ee
Equation (\ref{ex3}) now reads
\beaa
\Psi^0_+=\Psi^0_-\exp(\mathrm{i}F(\phi,-\lambda z+x)).
\eeaa
The solution to this equation looks like
\beaa
\Psi^0=\lambda\exp\left(\frac{1}{2\pi}
\int_\gamma\frac{\d \l'}{\l-\l'}F(\phi(\l'),-\lambda' z+x)
\right).
\eeaa
Considering the expansion of this expression in $\lambda$, we obtain
the equations
\bea
&&
\frac{1}{2\pi}
\int_\gamma F(\phi(\l),-\lambda z+x){\d \l}=0,
\\&&
\frac{1}{2\pi\text{i}}
\int_\gamma \l F(\phi(\l),-\lambda z+x){\d \l}=f.
\eea
Taking into account expression (\ref{phi}), we come to the conclusion
that these equations define the functions $u$, $f$ as implicit functions.
Solution to Dunajski equation is then defined by the relations
$$\Th_x=-u,\; \Th_y=zf.$$
Thus we have obtained a solution to Dunajski equation, depending on
arbitrary function of two variables, in terms of implicit functions.

Functional dependence on the function of two variables indicates
that the solution we have constructed correspons to some
(2+1)-dimensional reduction of Dunajski equation. It is possible
to find the reduced equations explicitely, using the fact that linear
equations (\ref{Lax2}) have  analytic (polynomial) 
solutions $\phi$ and $-\lambda z+x$.
Substituting these solutions to linear problems (\ref{Lax2}) and using 
relations (\ref{uvtheta}),
we obtain
\bea
&&
(\p_w-\Th_{xy}\p_y+\Th_{yy} \p_x)(2wf-\Th_y)+yf_y=0,
\label{firstred}
\\&&
(\p_z+\Th_{xx}\p_y-\Th_{xy} \p_x)(2wf-\Th_y)-yf_x=0,\nn
\\&&
zf=\Theta_y.\nn
\eea
\section{Dunajski equation hierarchy}
The framework developed here is closely connected with the framework of
of hyper-K\"ahler hierarchy developed by
Takasaki \cite{Takasaki,Takasaki1},
see also \cite{heav, HEred}. Though there are some essential differences
(the volume form is used instead of symplectic form, the spectral variable is included
to the form), the technique and ideas of the proofs are very similar. We should 
also mention an integrable generalization of the first heavenly equation
proposed by Park \cite{Park} and studied by Takasaki \cite{Takasaki0}, 
connected with volume-preserving diffeomorphisms. However, the hierarchy for
this model was not considered. Later a kind of volume preserving hierarchy was 
introduced in \cite{Guha}, but the hierarchy described in our work doesn't
coincide with it.

To define Dunajski equation hierarchy, we consider three formal
series, depending on two infinite sets of
additional variables (`times')
\bea
&&
\Psi^0=\lambda+\sum_{n=1}^\infty \Psi^0_n(\mathbf{t}^1,\mathbf{t}^2)\l^{-n},
\label{form0}
\\&&
\Psi^1=\sum_{n=0}^\infty t^1_n (\Psi^0)^{n}+
\sum_{n=1}^\infty \Psi^1_n(\mathbf{t}^1,\mathbf{t}^2)(\Psi^0)^{-n}
\label{form1}
\\&&
\Psi^2=\sum_{n=0}^\infty t^2_n (\Psi^0)^{n}+
\sum_{n=1}^\infty \Psi^2_n(\mathbf{t}^1,\mathbf{t}^2)(\Psi^0)^{-n},
\label{form2}
\eea
where $\mathbf{t}^1=(t^1_0,\dots,t^1_n,\dots)$, $\mathbf{t}^2=(t^2_0,\dots,t^2_n,\dots)$.
We denote $x=t^1_0$, $y=t^2_0$,
$
\mathbf{\Psi}=
\left(
\begin{array}{c}
\Psi^0\\
\Psi^1\\
\Psi^2
\end{array}
\right),
$
$\partial^1_n=\frac{\partial}{\partial t^1_n}$,
$\partial^2_n=\frac{\partial}{\partial t^2_n}$ and
introduce  the projectors $(\sum_{-\infty}^{\infty}u_n \l^n)_+
=\sum_{n=0}^{\infty}u_n \l^n$,
$(\sum_{-\infty}^{\infty}u_n \l^n)_-=\sum_{-\infty}^{n=-1}u_n \l^n$.

Dunajski equation hierarchy is defined by the relation
\be
(\d \Psi^0\wedge \d \Psi^1\wedge \d \Psi^2)_-=0,
\label{analyticity0}
\ee
where the differential includes both times and a
spectral variable,
\beaa
\d f=\sum_{n=0}^{\infty}\partial^1_n f \d t^1_n +
\sum_{n=0}^{\infty}\partial^2_n f \d t^2_n
+ \partial_\l f \d \l.
\eeaa
This is a crucial difference with the heavenly equation hierarchy,
where only the times are taken into account.
Relation (\ref{analyticity0})
plays a role similar to the role of the famous Hirota bilinear identity for KP hierarchy.
This relation is equivalent to the Lax-Sato form of
Dunajski equation hierarchy.
\begin{theorem}
Relation
(\ref{analyticity0})
is equivalent to the set of equations
\bea
&&
\partial^1_n\mathbf{\Psi}=\sum_{i=0,1,2}(J^{-1}_{1i} (\Psi^0)^n)_+
{\partial_i}\mathbf{\Psi},
\label{Dun1}
\\
&&
\partial^2_n\mathbf{\Psi}= \sum_{i=0,1,2}(J^{-1}_{2i} (\Psi^0)^n)_+
{\partial_i}\mathbf{\Psi},
\label{Dun2}
\\
&&
\det J=1,
\label{volume}
\eea
where
\be
J=
\begin{pmatrix}
\Psi^0_\l&\Psi^1_\l&\Psi^2_\l\\
\Psi^0_x&\Psi^1_x&\Psi^2_x\\
\Psi^0_y&\Psi^1_y&\Psi^2_y
\end{pmatrix},
\label{J}
\ee
$\partial_0=\partial_\lambda$, $\partial_1=\partial_x$, $\partial_2=\partial_y$.
\label{formDun}
\end{theorem}
The proof of (\ref{analyticity0}) $\Rightarrow$
hierarchy (\ref{Dun1},\ref{Dun2},\ref{volume}) is based on the following
statement.
\begin{lemma}
Given identity (\ref{analyticity0}), for arbitrary first order operator $\hat U$,
$$
\hat U\mathbf{\Psi} =\left(\sum_i (u^1_i(\lambda,\mathbf{t}^1,\mathbf{t}^2)\partial^1_i
+u^2_i(\lambda,\mathbf{t}^1,\mathbf{t}^2)\partial^2_i)+
u^0(\lambda,\mathbf{t}^1,\mathbf{t}^2)\partial_\lambda \right)\mathbf{\Psi}
$$
with `plus' coefficients ($(u^1_i)_-=(u^2_i)_-=u^0_-=0$), 
the condition $(\hat U\mathbf{\Psi})_+=0$
(for $\Psi^1$ and $\Psi^2$ modulo the derivatives of $\Psi^0$)
implies that $\hat
U\mathbf{\Psi}=0$.
\label{operator}
\end{lemma}
\textbf{Proof}
First, relation (\ref{analyticity0}) implies that
$$
(\det J)_-=0,
$$
and, using expansions (\ref{form0}), (\ref{form1}), (\ref{form2}), we get
$$
\det J=(\det J)_+=1.
$$
Then, using relation (\ref{analyticity0}) we obtain
(we use $|A|$ for $\det A$) 
\beaa
\begin{vmatrix}
\hat U \Psi^0 &\hat U \Psi^1 &\hat U \Psi^2 \\
\Psi^0_x&\Psi^1_x&\Psi^2_x\\
\Psi^0_y&\Psi^1_y&\Psi^2_y
\end{vmatrix}_-=
\begin{vmatrix}
\Psi^0_\l&\Psi^1_\l&\Psi^2_\l\\
\hat U\Psi^0&\hat U\Psi^1&\hat U\Psi^2\\
\Psi^0_y&\Psi^1_y&\Psi^2_y
\end{vmatrix}_-=
\begin{vmatrix}
\Psi^0_\l&\Psi^1_\l&\Psi^2_\l\\
\Psi^0_x&\Psi^1_x&\Psi^2_x\\
\hat U\Psi^0&\hat U\Psi^1&\hat U\Psi^2
\end{vmatrix}_-
=0
\eeaa
On the other hand, condition $(\hat U\mathbf{\Psi})_+=0$, taking into account
expansions (\ref{form0}), (\ref{form1}), (\ref{form2}), implies that
\beaa
\begin{vmatrix}
\hat U \Psi^0 &\hat U \Psi^1 &\hat U \Psi^2 \\
\Psi^0_x&\Psi^1_x&\Psi^2_x\\
\Psi^0_y&\Psi^1_y&\Psi^2_y
\end{vmatrix}_+=
\begin{vmatrix}
\Psi^0_\l&\Psi^1_\l&\Psi^2_\l\\
\hat U\Psi^0&\hat U\Psi^1&\hat U\Psi^2\\
\Psi^0_y&\Psi^1_y&\Psi^2_y
\end{vmatrix}_+=
\begin{vmatrix}
\Psi^0_\l&\Psi^1_\l&\Psi^2_\l\\
\Psi^0_x&\Psi^1_x&\Psi^2_x\\
\hat U\Psi^0&\hat U\Psi^1&\hat U\Psi^2
\end{vmatrix}_+
=0
\eeaa
Thus, we come to the conclusion that
\beaa
\begin{vmatrix}
\hat U \Psi^0 &\hat U \Psi^1 &\hat U \Psi^2 \\
\Psi^0_x&\Psi^1_x&\Psi^2_x\\
\Psi^0_y&\Psi^1_y&\Psi^2_y
\end{vmatrix}=
\begin{vmatrix}
\Psi^0_\l&\Psi^1_\l&\Psi^2_\l\\
\hat U\Psi^0&\hat U\Psi^1&\hat U\Psi^2\\
\Psi^0_y&\Psi^1_y&\Psi^2_y
\end{vmatrix}=
\begin{vmatrix}
\Psi^0_\l&\Psi^1_\l&\Psi^2_\l\\
\Psi^0_x&\Psi^1_x&\Psi^2_x\\
\hat U\Psi^0&\hat U\Psi^1&\hat U\Psi^2
\end{vmatrix}
=0.
\eeaa
This set of relations represents a linear system for $\hat U \mathbf{\Psi}$,
\beaa
J^{-1}_\text{tr}\hat U \mathbf{\Psi}=0,
\eeaa
and taking into account that $\det J=1$, the only solution to it is $\hat U\mathbf{\Psi}=0$.
\hfill$\square$\\
\\
The proof of the statement (\ref{analyticity0}) $\Rightarrow$ hierarchy 
(\ref{Dun1},\ref{Dun2},\ref{volume}) is then
straightforward, using simple relations
\beaa
\sum_{i=0,1,2}(J^{-1}_{1i} (\Psi^0)^n)
{\partial_i}\Psi^k=\delta_{1k}(\Psi^0)^n,\\
\sum_{i=0,1,2}(J^{-1}_{2i} (\Psi^0)^n)
{\partial_i}\Psi^k=\delta_{2k}(\Psi^0)^n.\\
\eeaa

The statement (\ref{Dun1},\ref{Dun2},\ref{volume}) $\Rightarrow$ 
(\ref{analyticity0}) directly follows
from the relation
\begin{lemma}
\be
\begin{vmatrix}
\partial_{\tau_0}\Psi^0&\partial_{\tau_0}\Psi^1&\partial_{\tau_0}\Psi^2\\
\partial_{\tau_1}\Psi^0&\partial_{\tau_1}\Psi^1&\partial_{\tau_1}\Psi^2\\
\partial_{\tau_2}\Psi^0&\partial_{\tau_2}\Psi^1&\partial_{\tau_2}\Psi^2
\end{vmatrix}
=
\begin{vmatrix}
V_{\tau_0+}^0&V_{\tau_0+}^1&V_{\tau_0+}^2\\
V_{\tau_1+}^0&V_{\tau_1+}^1&V_{\tau_1+}^2\\
V_{\tau_2+}^0&V_{\tau_2+}^1&V_{\tau_2+}^2
\end{vmatrix}
\ee
where $\tau_0$, $\tau_1$, $\tau_2$ is an arbitrary set of three times of the hierarchy
(\ref{Dun1},\ref{Dun2},\ref{volume}), and $V_{\tau+}^i$ are the 
coefficients of corresponding vector fields
given by the r.h.s. of equations (\ref{Dun1},\ref{Dun2}),
$$
\partial_{\tau_j} \mathbf{\Psi}=\sum_{i=0,1,2}V_{\tau_j+}^i
{\partial_i}\mathbf{\Psi}.
$$
\end{lemma}
\textbf{Proof}
\beaa
\det (\partial_{\tau_j}\Psi^k)= \det(\sum_{i=0,1,2}V_{\tau_j+}^i
{\partial_i}\Psi^k)=\det(V_{\tau_j+}^i) \cdot\det J=\det(V_{\tau_j+}^i)
\eeaa
\hfill$\square$\\
To complete the picture, we will also prove compatibility of equations of the hierarchy
(\ref{Dun1},\ref{Dun2},\ref{volume}).
\begin{theorem}
The flows of the
hierarchy (\ref{Dun1},\ref{Dun2}) commute and the condition $\det J=1$(\ref{volume}) is
preserved by the dynamics.
\end{theorem}
\textbf{Proof} Let us consider, e.g.,
the following flows of the hierarchy:
\bea
&&
\partial^1_n\mathbf{\Psi}=\sum_{i=0,1,2}(J^{-1}_{1i} (\Psi^0)^n)_+
{\partial_i}\mathbf{\Psi}=\hat{V}^1_{n+}\mathbf{\Psi},
\label{Dun1n}
\\
&&
\partial^2_m\mathbf{\Psi}= \sum_{i=0,1,2}(J^{-1}_{2i} (\Psi^0)^m)_+
{\partial_i}\mathbf{\Psi}=\hat{V}^2_{m+}\mathbf{\Psi},
\label{Dun2m}
\eea
where $\hat{V}^1_{n}=\sum_{i}(J^{-1}_{1i} (\Psi^0)^n)
{\partial_i}$, $\hat{V}^2_{m}=\sum_{i}(J^{-1}_{2i} (\Psi^0)^m)
{\partial_i}$ are vector fields, possessing the property
\bea
&&\hat{V}^1_{n}\Psi^k=\delta_{1k}(\Psi^0)^n,
\label{id1}
\\
&&\hat{V}^2_{m}\Psi^k=\delta_{2k}(\Psi^0)^m.
\label{id2}
\eea
The condition of compatibility of the flows (\ref{Dun1n}), (\ref{Dun2m}) is
\be
\partial^2_m(\hat{V}^1_{n+}\mathbf{\Psi})=\partial^1_n(\hat{V}^2_{m+}\mathbf{\Psi}).
\label{compat0}
\ee
First, in a standard way,
$$
\partial^2_m(\hat{V}^1_{n+}\mathbf{\Psi})-\partial^1_n(\hat{V}^2_{m+}\mathbf{\Psi})=
\left(\partial^2_m\hat{V}^1_{n+}-\partial^1_n\hat{V}^2_{m+}+[\hat{V}^1_{n+},
\hat{V}^2_{m+}]\right)
\mathbf{\Psi}.
$$
Relation $\hat{U}\mathbf{\Psi}=0$ for some vector field $\hat{U}$
may be considered as a homogeneous linear system for the coefficients of the
vector
field $\hat{U}$ with the matrix $J$ (\ref{volume}). 
Thus $\hat{U}\mathbf{\Psi}=0\Rightarrow \hat{U}=0$,
and condition (\ref{compat0}) is equivalent to the usual form of compatibility condition,
\be
\partial^2_m\hat{V}^1_{n+}-\partial^1_n\hat{V}^2_{m+}+[\hat{V}^1_{n+},\hat{V}^2_{m+}]=0.
\label{compat}
\ee

On the other hand, using (\ref{id1},\ref{id2}), we get
$$
\partial^2_m(\hat{V}^1_{n+}\mathbf{\Psi})-\partial^1_n(\hat{V}^2_{m+}\mathbf{\Psi})=-
\left(\partial^2_m\hat{V}^1_{n-}-
\partial^1_n\hat{V}^2_{m-}-[\hat{V}^1_{n-},\hat{V}^2_{m-}]\right)
\mathbf{\Psi}
$$
Then
\beaa
&&\hat{W}\mathbf{\Psi}=(\hat{W}_+ + \hat{W}_-) \mathbf{\Psi}=0,
\\
&&
\hat{W}_+=\left(\partial^2_m\hat{V}^1_{n+}-\partial^1_n\hat{V}^2_{m+}+
[\hat{V}^1_{n+},\hat{V}^2_{m+}]\right)
\\
&&
\hat{W}_-=\left(\partial^2_m\hat{V}^1_{n-}-\partial^1_n\hat{V}^2_{m-}-
[\hat{V}^1_{n-},\hat{V}^2_{m-}]\right),
\eeaa
and we come to the conclusion that $\hat{W}=0$, so, evidently, $\hat{W}_+=\hat{W}_-=0$,
that proves compatibility condition (\ref{compat}).

The conservation of volume (\ref{volume}) by some flow
$\partial_{\tau} \mathbf{\Psi}=\sum_{i}V_{\tau+}^i
{\partial_i}\mathbf{\Psi}$
is equivalent to the zero divergence condition 
\be
\sum_i \partial_i V_{\tau+}^i=0, 
\label{divergence}
\ee
which 
can be checked directly
starting from the definition of the hierarchy.
Indeed,
\beaa
&&
J^{-1}_{1i}=-\sum_{j,k} \epsilon_{ijk}\partial_j\Psi^0 \partial_k \Psi^2,
\\
&&
J^{-1}_{2i}=\sum_{j,k}\epsilon_{ijk} \partial_j\Psi^0 \partial_k \Psi^1,
\eeaa
and calculation of divergence (\ref{divergence}) for any of the flows
(\ref{Dun1},\ref{Dun2}) leads to appearence of symmetry for a pair of indices
under summation with completely antisymmetric symbol $\epsilon_{ijk}$, giving
zero as the result.
\hfill$\square$\\
\\
In a more explicit form, Dunajski equation hierarchy
(\ref{Dun1}, \ref{Dun2}) can be written as
\bea
\partial^1_n\mathbf{\Psi}=
+\left(
(\Psi^0)^{n}
\begin{vmatrix}
\Psi^0_\l & \Psi^2_\l\\
\Psi^0_y & \Psi^2_y
\end{vmatrix}
\right)_+\partial_x \mathbf{\Psi}-
\left(
(\Psi^0)^{n}
\begin{vmatrix}
\Psi^0_\l & \Psi^2_\l\\
\Psi^0_x & \Psi^2_x
\end{vmatrix}
\right)_+\partial_y \mathbf{\Psi}-
\nn\\
\left(
(\Psi^0)^{n}
\begin{vmatrix}
\Psi^0_x & \Psi^2_x\\
\Psi^0_y & \Psi^2_y
\end{vmatrix}
\right)_+\partial_\l \mathbf{\Psi}
,
\label{Dun11}
\eea
\bea
\partial^2_n\mathbf{\Psi}=
-\left(
(\Psi^0)^{n}
\begin{vmatrix}
\Psi^0_\l & \Psi^1_\l\\
\Psi^0_y & \Psi^1_y
\end{vmatrix}
\right)_+\partial_x \mathbf{\Psi}+
\left(
(\Psi^0)^{n}
\begin{vmatrix}
\Psi^0_\l & \Psi^1_\l\\
\Psi^0_x & \Psi^1_x
\end{vmatrix}
\right)_+\partial_y \mathbf{\Psi}+
\nn\\
\left(
(\Psi^0)^{n}
\begin{vmatrix}
\Psi^0_x & \Psi^1_x\\
\Psi^0_y & \Psi^1_y
\end{vmatrix}
\right)_+\partial_\l \mathbf{\Psi}
\label{Dun21}
\eea
(plus equation (\ref{volume})).
It is easy to check that for $\Psi^0=\lambda$ Dunajski equation hierarchy reduces
to heavenly equation hierarchy
\cite{Takasaki,Takasaki1}, while for $\Psi^2=y$ it reduces to dispersionless
KP hierarchy. The first two flows of the hierarchy (\ref{Dun11}), (\ref{Dun21})
read
\beaa
&&\p_1^1\mathbf{\Psi}=(u_{y}\p_x-u_{x}\p_y+\l\p_x-f_{x}\p_{\l})\mathbf{\Psi},
\\
&&\p_1^2\mathbf{\Psi}=
(v_{x}\p_y-v_{y}\p_x +\l\p_y-f_{y}\p_{\l})\mathbf{\Psi},
\eeaa
which yields Lax pair (\ref{Lax2}) after the identification
$z=-t^1_1$, $w=t^2_1$.
Here
$$
u=\Psi^2_1,\quad v=\Psi^1_1, \quad f=\Psi^0_1,
$$
and condition $\det J=1$ (\ref{volume}) implies that $u_y+v_x=0$.

\subsection{Related hierarchies}
Formula (\ref{volume}) defines  a reduction for equations
(\ref{Dun1}),  (\ref{Dun2}).
The general hierarchy in the unreduced case is given by equations
(\ref{Dun1}), (\ref{Dun2}),
\bea
&&
\partial^1_n\mathbf{\Psi}=\sum_{i=0,1,2}(J^{-1}_{1i} (\Psi^0)^n)_+
{\partial_i}\mathbf{\Psi},
\label{Dun1gen}
\\
&&
\partial^2_n\mathbf{\Psi}= \sum_{i=0,1,2}(J^{-1}_{2i} (\Psi^0)^n)_+
{\partial_i}\mathbf{\Psi},
\label{Dun2gen}
\eea
and the analogue of relation (\ref{analyticity0}) is
\be
((\det J)^{-1}\d \Psi^0\wedge \d \Psi^1\wedge \d \Psi^2)_-=0.
\label{analyticity00}
\ee
For the unreduced hierarchy the propositions formulated above and their
proofs are completely analogous. The hierarchy may also be considered for arbitrary 
number of components $\Psi^0$, $\Psi^i$.

Two-component case of relation (\ref{analyticity00})
\bea
((\det J)^{-1}\d \Psi^0\wedge \d \Psi^1)_-=0
\label{twocomp}
\eea
and corresponding equations (\ref{Dun1gen})
define a hierarchy for the system
introduced in \cite{MS2} (dispersionless KP minus area conservation).
If $\det J=1$, relation (\ref{twocomp}) defines dispersionless KP hierarchy.

A special subclass of the hierarchies of the type (\ref{analyticity00}),
(\ref{analyticity0}) is singled out by the condition $\Psi^0=\lambda$. In
this case (\ref{analyticity0}) is transformed to the relation
\beaa
(\d \lambda \wedge \d \Psi^1\wedge \d \Psi^2)_-=0,
\eeaa
which is equivalent to
\bea
(\tilde\d \Psi^1\wedge \tilde\d \Psi^2)_-=0,
\label{heav}
\eea
where the differential $\tilde\d$ takes into account only times
(and not a spectral variable). As it is known, relation (\ref{heav})
defines Plebanski second heavenly equation hierarchy
\cite{Takasaki,Takasaki1} (see also \cite{heav,HEred}).
A two-component case of (\ref{analyticity00}) (relation (\ref{twocomp})) 
under the condition
$\Psi^0=\lambda$ reduces to
$$
((\Psi^1_x)^{-1}\tilde\d \Psi^1)_-=0.
$$
This relation defines a hierarchy considered in \cite{ASha}, 
corresponding equations  (\ref{Dun1gen}) define a system of
(positive) flows of this hierarchy.

\section*{Acknowledgment}
The authors were partially supported by Russian Foundation for
Basic Research under grant no. 06-01-90840 and grant no. 06.01 CRF
CSTD AS Moldova. LVB and SVM were also supported in part by RFBR grants 
07-01-00446 and 06-01-92053 and LVB was supported in part by RFBR grant
06-01-89507.


\end{document}